\documentclass[pra,twocolumn]{revtex4}
\usepackage{graphicx}
\usepackage{enumerate}
\usepackage{graphicx}
\usepackage{amsmath}
\usepackage{amsfonts}
\usepackage{amssymb}
\usepackage{color}
\usepackage{bm}
\newcommand{\ket}[1]{\left| #1 \right\rangle}
\newcommand{\bra}[1]{\left\langle #1 \right|}

\begin{document}
\title{Transitionless quantum driving in spin echo}
\author{Anton Gregefalk}
\affiliation{Department of Physics and Astronomy, Uppsala University, Box 516,
Se-751 20 Uppsala, Sweden}
\author{Erik Sj\"oqvist}
\email{erik.sjoqvist@physics.uu.se}
\affiliation{Department of Physics and Astronomy, Uppsala University, Box 516,
Se-751 20 Uppsala, Sweden}
\date{\today}
\begin{abstract}
Spin echo can be used to refocus random dynamical phases caused by inhomogeneities 
in control fields and thereby retain the purity of a spatial distribution of quantum spins. 
This technique for accurate spin control is an essential ingredient in many applications, 
such as nuclear magnetic resonance, magnetic resonance imaging, and quantum 
information processing. Here, we show how all the elements of a spin echo sequence 
can be performed at high speed by means of transitionsless quantum driving. 
This technique promises accurate control of rapid quantum spin evolution. We apply the 
scheme to universal nonadiabatic geometric single- and two-qubit gates in a nuclear 
magnetic resonance setting.  
\end{abstract}
\maketitle

\section{Introduction}
Techniques for accurate spin control is an essential ingredient in physics and chemistry. 
One such technique is spin echo \cite{hahn50,carr54}, in which a spin is taken along a 
path twice, where the second path exactly retraces the first one but in opposite direction, 
and where the paths are surrounded by short $\pi$ pulses. In this way, it becomes possible 
to refocus random dynamical phases caused by spatial inhomogeneities in the 
control fields and thereby retain the purity of the spin ensemble. 
  
More generally, the sensitive nature of the quantum regime makes the system prone to 
errors, where tiny disturbances can render the quantum nature of the system to disappear. 
This becomes in particular challenging in quantum information processing (QIP), as loss of 
coherence removes the advantages of using quantum degrees of freedom as information 
carriers. Therefore, various refocusing techniques have been used in QIP, such as, e.g., in 
the implementation of robust geometric gates in adiabatic cyclic evolution, in which the use 
of Berry phases \cite{berry84} removes the dependence of experimental details 
\cite{jones00,ekert00}. 

The use of adiabatic evolution in spin echo prolong the exposure to  errors of 
quantum-mechanical origin, such as environment-induced decoherence and decay. This calls 
for a decrease of the run time of the adiabatic paths. Here, we address this problem by 
combining the idea of spin echo with transitionsless quantum driving (TQD) \cite{berry09} to shorten 
the run time and to suppress nonadiabatic transitions. This scheme opens up for precise 
control of high-speed evolution of quantum spins, which may help to improve the accuracy 
of various technological applications, such as nuclear magnetic resonance (NMR) \cite{darbeau06}, 
magnetic resonance imaging (MRI) \cite{jung13}, and QIP \cite{vandersypen05}.  

\section{The scheme}
\begin{figure}[htb]
\centering
\includegraphics[width=0.16\textwidth]{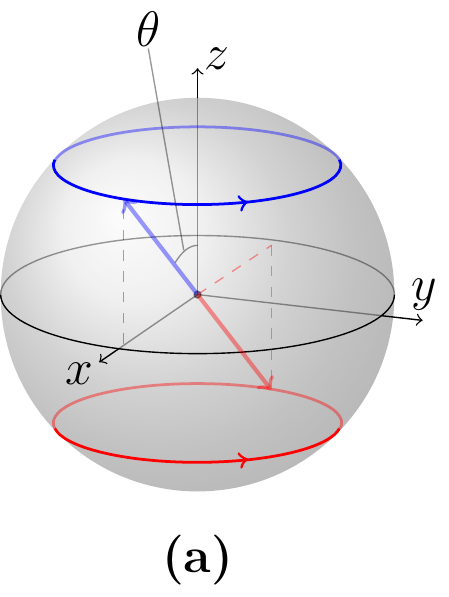} \hskip 1 cm
\includegraphics[width=0.16\textwidth]{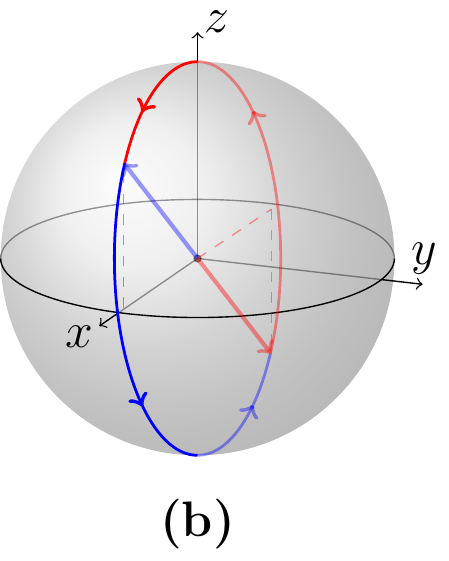} \hskip 1 cm
\includegraphics[width=0.16\textwidth]{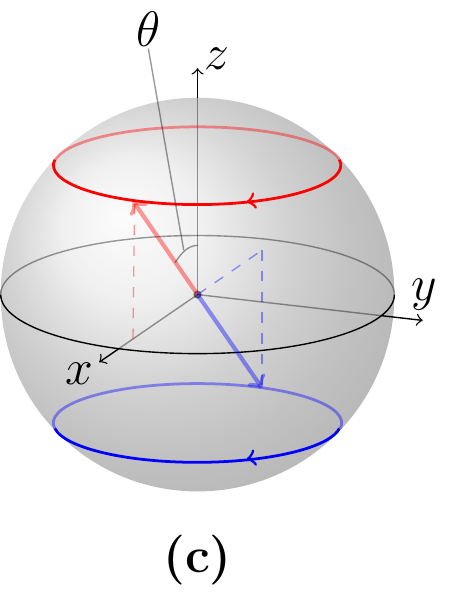} \hskip 1 cm
\includegraphics[width=0.16\textwidth]{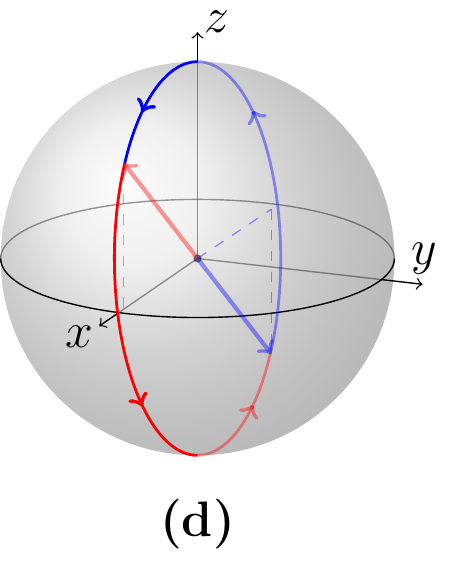} 
\caption{The basic spin echo sequence. (a) and (c) are the two closed adiabatic 
spin loops traced in opposite directions. (b) and (d) are short $\pi$ transformations, 
half-circle rotations of the spin around the $y$ axis.}
\label{fig:spinecho}
\end{figure}

We consider a spin echo scheme based on closed loops of adiabatic control parameters. 
Our basic physical setting is a spin qubit in an NMR system in which a radio frequency (rf) 
magnetic field in the $xy$ plane is added to a bias magnetic field in the $z$ direction. The 
latter is detuned in a rotating frame by the frequency of the rf field. By sweeping the rf frequency 
not all away to resonance, the initial spin can be 
made to point at an arbitrary angle $\theta$ to the rotational $z$ axis. The spin is thereafter 
taken around a cone-shaped loop by rotating the rf field slowly with angular frequency $\omega$ 
around the $z$ axis \cite{jones00,ekert00}. This implies that the two orthogonal spin states 
(we consider spin $s=\frac{1}{2}$) pick up Berry phases 
$\beta_{\pm} = \mp \pi (1-\cos \theta)$. A second loop surrounded by short $\pi$ pulses 
exactly retraces the first loop but in opposite direction. The basic adiabatic sequence is 
shown in Fig.~\ref{fig:spinecho}.  

We now demonstrate how TQD can be used to implement a modified scheme, in which 
all components of the spin echo sequence can be performed at high speed. While our 
approach is adapted to NMR, a similar technique has been used to measure the Berry 
phase at high speed in a superconducting phase qubit \cite{zhang17}. 

Consider a spin-$\frac{1}{2}$ in a time dependent magnetic field ${\bf B}_0 = B_0 
(\sin \theta \cos \omega t, \sin \theta \sin \omega t, \cos \theta)$, defining the 
`root' Hamiltonian 
\begin{eqnarray}
H_0(t) = \gamma {\bf B}_0(t)\cdot {\bm S} ,
\end{eqnarray}
of the TQD setting. Here, $\gamma$ is the gyromagnetic ratio and ${\bm S} = 
\frac{1}{2} \hbar \bm{\sigma}$ with $\bm{\sigma}=(\sigma_x,\sigma_y,\sigma_z)$ 
the standard Pauli operators. Nonadiabatic transitions between 
the instantaneous eigenstates $\ket{\phi_{\pm} (t)}$ of $H_0(t)$ are completely 
suppressed by adding a correction term, yielding the TQD Hamiltonian \cite{berry09}
\begin{eqnarray}
H(t) = \left[ \gamma{\bf B}_0(t)+{\bf b}_0(t) \times \partial_t {\bf b}_0(t) \right] \cdot{\bm S} =  
\gamma {\bf B}(t) \cdot {\bm S}
\end{eqnarray}
with ${\bf b}_0(t) = {\bf B}_0(t) / \left| {\bf B}_0(t) \right|$ the direction of the magnetic field. 
In this way, the spin prepared at $t=0$ in an eigenstate $\ket{\phi_{\pm} (0)}$ of $H_0 (0)$ 
exactly monitors the instantaneous eigenstate $\ket{\phi_{\pm} (t)}$ of $H_0 (t)$. In other 
words, the exact state at $t\geq 0$ reads 
\begin{eqnarray}
\ket{\psi (t)} = e^{i f(t)} \ket{\phi_{\pm} (t)}
\end{eqnarray}
no matter how fast $H_0(t)$ varies with time $t$. 

By using ${\bf b}_0(t)$, one finds the explicit form of the TQD field \cite{berry09}:  
\begin{eqnarray}
\gamma {\bf B}(t) & = & \left( \omega_0-\omega\cos\theta \right) \sin\theta \left[ {\bf e}_x
\cos (\omega t) + {\bf e}_y \sin (\omega t) \right] 
\nonumber \\
 & & + \left( \omega_0 \cos\theta + \omega \sin^2\theta \right) \hat{e}_z , 
 \label{eq:stafield}
\end{eqnarray}
where from now on the field strength $B_0(t) = B_0$ is taken to be constant and we have 
put $\omega_0 = \gamma B_0$. This leads to the TQD Hamiltonian 
\begin{eqnarray} 
H(t) & = & \frac{1}{2} \hbar \left[ \left( \omega_0 - \omega \cos \theta \right) \sin \theta
\right] \left[ \cos (\omega t) \sigma_x +\sin (\omega t) \sigma_y \right] 
\nonumber \\ 
 & & + \frac{1}{2} \hbar \left( \omega_0 \cos\theta + \omega \sin^2 \theta 
\right) \sigma_z . 
\label{eq:hamstart}
\end{eqnarray}

The refocusing sequence shown in Fig.~\ref{fig:spinecho} is 
\begin{eqnarray}
C\to \pi \to \bar{C} \to \pi,
\end{eqnarray}
where $C$ is the loop traced by the eigenstates $\ket{\phi_{\pm} (t)}$ of the Hamiltonian 
$H_0(t)$ exactly driven by $H(t)$, and $\bar{C}(\omega) = C(-\omega)$ is the same loop 
traced backwards. The $\pi$ transformation is enacted after each loop and described by 
the Hamiltonian 
\begin{eqnarray}
H(t) = \frac{1}{2} \hbar \omega_\pi \sigma_y,
\end{eqnarray}
by applying the magnetic field  $\gamma {\bf B}_\pi(t) = (0,\omega_\pi,0)$. 
The sequence can now be described by a Hamiltonian divided up as 
\begin{eqnarray}
\label{eq:hamseq}
H(t) = \left\{\begin{array}{ll}
H^{(a)}(t) , & \ 0\leq t \leq \frac{2\pi}{\omega}, \\
H^{(b)}(t) , & \ t_1\leq t\leq t_1 +\frac{\pi}{\omega_\pi}, \\
H^{(c)}(t) , & \ t_2\leq t \leq t_2+\frac{2\pi}{\omega}, \\
H^{(d)}(t) , & \ t_3\leq t\leq t_3 +\frac{\pi}{\omega_\pi}\equiv \tau,
\end{array}\right.
\end{eqnarray}
where $t_1 > \frac{2\pi}{\omega}$, $t_2 > t_1+\frac{\pi}{\omega_\pi}$, and 
$t_3>t_2+\frac{2\pi}{\omega}$ (the Hamiltonian is assumed to vanish during the intermediate 
time intervals $[\frac{2\pi}{\omega},t_1]$, $[t_1 +\frac{\pi}{\omega_\pi},t_2]$, and 
$[t_2+\frac{2\pi}{\omega},t_3]$). Here, the TQD Hamiltonians $H^{(a)}(t)$ and $H^{(c)}(t)$ 
exactly implement the two spin loops $C$ and $\bar{C}$, respectively, 
while $H^{(b)}(t)$ and $H^{(d)}(t)$ correspond to the two $\pi$ pulses. 

\begin{figure}[htb]
\centering
\includegraphics[width=0.35\textwidth]{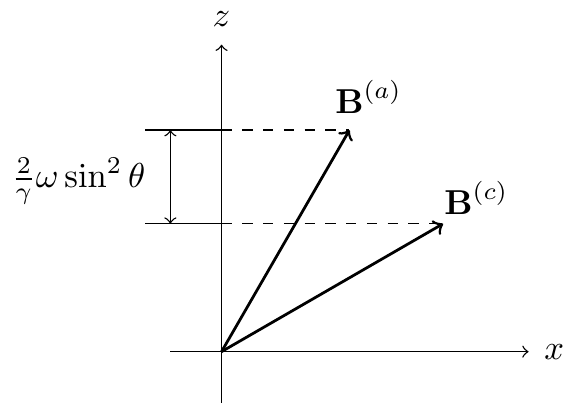} 
\caption{The difference in magnetic field for the first and second loop, in the $xz$ plane.}
\label{fig:magfield}
\end{figure}

In order for the second loop to retrace the first one, the corresponding magnetic 
fields ${\bf B}^{(c)}$ and ${\bf B}^{(a)}$ must have different opening angles to the rotational 
axis. Indeed, from Eq.~(\ref{eq:stafield}), we find the relative vector $\gamma \Delta {\bf B} = 
\gamma \left( {\bf B}^{(a)}-{\bf B}^{(c)} \right) = 2\sin\theta(-\omega\cos\theta\cos\omega t, 
\omega_0\sin\omega t,\omega\sin\theta)$, which is clearly nonzero and scales linearly with 
$\omega$, meaning that the speed of the evolution dictates the difference in applied 
magnetic fields. This is a direct consequence of the TQD since the $\omega$ 
factor arose from the correction term, and since the adiabatic regime is entered when 
$\left| \omega /\omega_0 \right| \ll 1$, in which the relative vector would be close to 
zero. The respective magnitudes do not differ because $\gamma \left| {\bf B}^{(a)} \right| = 
\gamma \left| {\bf B}^{(c)} \right| = \omega_0 \left[ 1+\left( \frac{\omega}{\omega_0} \sin\theta \right)^2 
\right]^{1/2}$ is unaffected by a sign change of $\omega$. The discussed aspects are illustrated 
in Fig.~\ref{fig:magfield}.

Worth stressing is that the states driven are not the eigenstates of $H(t)$ but of the root 
Hamiltonian $H_0(t)$. The TQD technique enables the evolution to exactly track the 
instantaneous eigenstates of $H_0(t)$, even though the evolution is performed at 
high speed. A consequence of this is that the Berry phase is unaffected by the 
TQD, apart from being picked up faster. 

The dynamical phases are unaffected too by the TQD. This  can be seen by noting that 
the exact spin states take the form 
\begin{eqnarray}
\rho_{\pm} (t) = \frac{1}{2} \left( \hat{1} \pm {\bf b}_0 (t) \cdot \bm{\sigma} \right) , 
\end{eqnarray}
yielding  
\begin{eqnarray}
 & & {\rm Tr} \left\{ \left[ {\bf b}_0 (t) \times  \partial_t {\bf b}_0(t) \right] \cdot \bm{\sigma} 
\rho_{\pm} (t) \right\} 
\nonumber \\ 
 & = & \frac{1}{2} {\rm Tr} \left\{ \left[ {\bf b}_0 (t) \times  \partial_t 
{\bf b}_0 (t) \right] \cdot  \bm{\sigma} \right\} 
\nonumber \\ 
 & & \pm \frac{1}{2} {\rm Tr} \left\{ 
\left[ {\bf b}_0 (t) \times  \partial_t {\bf b}_0(t) \right] \cdot {\bf b}_0 (t) \hat{1} \right\} = 0 ,
\end{eqnarray}
where we have used that $\bm{\sigma}$ is traceless, $\bm{\sigma} \cdot \bm{\sigma} = \hat{1}$, 
and that the triple product $\left[ {\bf b}_0 (t) \times  \partial_t {\bf b}_0(t) \right] \cdot {\bf b}_0 (t)$ 
vanishes. In other words, the extra term in the TQD Hamiltonian causes no extra contribution 
to the spin energy, which implies that the dynamical phases are unaltered.  

Apart from the complete suppression of nonadiabatic corrections achieved by TQD, it 
should  be noted that the TQD spin echo scheme can be expected to have a similar kind 
of resilience to parameter noise and systematic errors as in standard adiabatic spin echo. 
The key advantage is instead the possibility to perform the spin echo sequence at high 
speed. In this way, the total run time can be made much shorter than that of spin relaxation 
and dephasing, which, e.g., in some QIP implementations 
may be as short as a few $\mu$s. This can be achieved by using fields that rotates at 
frequencies in the order of GHz or faster, which is feasible in typical NMR experiments. 

\section{Application: universal geometric gates}
TQD has been used to optimize NMR quantum information processing \cite{santos20} 
and to implement geometric quantum gates in nitrogen-vacancy (NV) centers 
\cite{liang16,kleissler18}. Motivated by these earlier results, we demonstrate that spin 
echo combined with TQD can be used for implementing a set of universal nonadiabatic 
geometric single- and two-qubit gates, providing means to speed up earlier geometric 
schemes \cite{jones00,ekert00} in NMR. 

Let us start with the single-qubit case. The root Hamiltonian $H_0 (t)$ is diagonalised by its 
instantaneous eigenvectors 
\begin{eqnarray}
\ket{\phi_0 (t)} & = & \cos\frac{\theta}{2}\ket{0}+\sin\frac{\theta}{2}e^{i\omega t}\ket{1} ,
\nonumber \\
\ket{\phi_1 (t)} & = & -\sin\frac{\theta}{2}\ket{0}+\cos\frac{\theta}{2}e^{i\omega t}\ket{1}  
\end{eqnarray}
with $\sigma_z \ket{p} = (1-2p) \ket{p}$, $p=0,1$, where we for notational convenience 
use the conventional qubit notation $\ket{0},\ket{1}$. Clearly, the initial eigenvectors 
$\ket{\phi_p (0)} \equiv \ket{\phi_p}$ are parametrised solely by the spherical angle $\theta$, 
and evolve exactly under the TQD Hamiltonian into the instantaneous eigenvectors 
$\ket{\phi_p (t)}$ up to phase factors. After completing a loop, each such 
phase factor comprises a dynamical ($\delta_p$) and a geometric ($\beta_p$) 
component, which, as shown above, are the same as those of adiabatic evolution 
driven by $H_0 (t)$ alone. Explicitly, one finds 
\begin{eqnarray}
\delta_p & = & (1-2p) \delta , 
\nonumber \\ 
\beta_p & = & (2p-1) \frac{1}{2} \Omega ,
\label{eq:phases}
\end{eqnarray}
where $\Omega$ is the solid angle enclosed by the loop. Note that while $\delta$ 
is independent of the orientation of the loop, $\Omega$ changes sign when the loop 
is reversed. Thus, $\tilde{\delta}_p = \delta_p$ and $\tilde{\beta}_p = 
-\beta_p$, which applied to the spin echo scheme results in  
\begin{eqnarray}
\ket{\phi_p} & \overset{C_p}{\to} & e^{i(\delta_p+\beta_p)}\ket{\phi_p} 
\overset{\pi}{\to}e^{i(\delta_p+\beta_p)}\ket{\phi_{p\oplus1}}
\nonumber \\ 
 & \overset{\bar{C}_p}{\to} & e^{i(\delta_p+\tilde{\delta}_{p \oplus 1}+\beta_p + 
\tilde{\beta}_{p\oplus 1})}\ket{\phi_{p\oplus1}} 
\nonumber \\ 
 & & = e^{i(\delta_p+\delta_{p\oplus 1}+\beta_p - 
\beta_{p\oplus 1})}\ket{\phi_{p\oplus1}} 
\nonumber \\ 
 & \overset{\pi}{\to} & 
e^{i(\delta_p+\delta_{p\oplus 1}+\beta_p - \beta_{p\oplus 1})}\ket{\phi_p},
\label{eq:spinecho}
\end{eqnarray}
where $\oplus$ is addition modulo 2. By combining Eqs.~(\ref{eq:phases}) and (\ref{eq:spinecho}), 
we see that the spin echo sequence cancels the dynamical phases and thereby results in the 
purely geometric gate $\ket{\phi_p} \to U(\Omega)\ket{\phi_p} = e^{i(2p-1)\Omega}\ket{\phi_p}$, 
i.e., 
\begin{eqnarray}
U(\Omega) = e^{-i\Omega}\ket{\phi_0} \bra{\phi_0}+e^{i\Omega}\ket{\phi_1} \bra{\phi_1}.
\end{eqnarray}
By rewriting this in terms of the computational basis $\{ \ket{0},\ket{1} \}$ via the expressions 
for the initial eigenstates $\ket{\phi_p}$, one obtains the unitary
\begin{eqnarray}
U(\theta,\Omega) =  e^{-i\Omega {\bf n}\cdot \bm{\sigma}} , 
\end{eqnarray}
where ${\bf n} = (\sin \theta , 0 ,\cos \theta)$. 

For $U(\theta,\Omega$) to be universal $\theta$ and $\Omega$ must be able to vary 
independently, but this is clearly not the case since $\left| \Omega \right| = 2\pi(1-\cos\theta)$, 
and so another control parameter $\vartheta$ is required. This can be obtained by rotating 
the TQD magnetic fields around some symmetry axis \cite{zhu02}, here, the $y$ axis, such that 
${\bf B}' = R_y (\vartheta-\theta){\bf B}$, in turn producing the rotated direction 
${\bf n}' = R_y(\vartheta-\theta){\bf n}$, i.e.,
\begin{eqnarray}
{\bf n}' & = & \begin{pmatrix}
\cos(\vartheta-\theta) & 0 & \sin(\vartheta-\theta)\\
 0 & 1 & 0\\
 -\sin(\vartheta-\theta) & 0 & \cos(\vartheta-\theta)
\end{pmatrix}
\begin{pmatrix}
\sin\theta \\ 
0 \\ 
\cos\theta
\end{pmatrix} 
\nonumber \\ 
 & = & \begin{pmatrix}
\sin\vartheta \\ 0\\ \cos\vartheta
\end{pmatrix}.
\end{eqnarray}
The result is the universal unitary 
\begin{eqnarray}
 & & U(\vartheta,\Omega) 
\nonumber \\ 
 & = & \begin{pmatrix}
\cos^2\frac{\vartheta}{2} e^{-i\Omega}+\sin^2\frac{\vartheta}{2} 
e^{i\Omega} & -i\sin\vartheta\sin\Omega \\
-i\sin\vartheta\sin\Omega & \sin^2\frac{\vartheta}{2} 
e^{-i\Omega}+\cos^2\frac{\vartheta}{2} e^{i\Omega}
\end{pmatrix} 
\nonumber \\ 
 & = & e^{-i\Omega {\bf n}'\cdot \bm{\sigma}}, 
\end{eqnarray}
where, by definition, $\vartheta$ and $\Omega$ can now be varied independently. 
Moreover, due to the spherical symmetry the Berry phases remain the same. This 
is because the same loops are traced, only in a rotated frame, and as such the solid 
angles remain the same.

For two unitaries $U(\vartheta_1,\Omega_1),U(\vartheta_2,\Omega_2)$ to contribute 
to universality it is necessary that $[U(\vartheta_1,\Omega_1),U(\vartheta_2,\Omega_2)]\neq 0$, 
which can be shown to equate to \cite{zhu02}
\begin{eqnarray}
\sin\Omega_1\sin\Omega_2\sin(\vartheta_1-\vartheta_2) \neq 0.
\end{eqnarray}
Simply choosing $\vartheta_1-\vartheta_2\neq n \pi$, where $n\in \mathbb{Z}$, 
takes care of the issue. By letting $\vartheta_1=0$, the unitary
\begin{eqnarray}
U(0,\Omega_1) = \begin{pmatrix}
  e^{-i\Omega_1}& 0\\
  0 & e^{i\Omega_1}
 \end{pmatrix},
\end{eqnarray}
is obtained which is equivalent to the $\phi$-gate $\ket{p} \to e^{ip2\Omega_1} \ket{p}$ up 
to a global phase $e^{-i\Omega_1}$. Now for the second gate, let $\vartheta_2=\pi/2$, then
\begin{eqnarray}
U \left( \frac{\pi}{2},\Omega_2 \right) = 
\begin{pmatrix}
\cos\Omega_2 & -i\sin\Omega_2\\
-i\sin\Omega_2 & \cos\Omega_2
\end{pmatrix}.
\end{eqnarray}
Furthermore, either set $\Omega_2 = \pi/2$ to obtain 
\begin{eqnarray}
U\left( \frac{\pi}{2},\frac{\pi}{2} \right) = 
\begin{pmatrix}
 0 & -i\\
 -i & 0
\end{pmatrix}.
\end{eqnarray}
being the equivalent to spin-flip up to the phase factor $-i$, or set $\Omega_2 = \pi/4$ 
to obtain
\begin{eqnarray}
U\left( \frac{\pi}{2},\frac{\pi}{4} \right) = \frac{1}{\sqrt{2}}\begin{pmatrix}
 1 & -i\\
 -i & 1
\end{pmatrix}.
\end{eqnarray}
which is an ``equal weighted superposition gate", being roughly equivalent to 
the Hadamard gate. 

We next extend the scheme to enact a geometric two-qubit gate by using the Ising 
interaction. The original Hamiltonian is 
\begin{eqnarray}
H_0(t) & = & \gamma_{\rm I} \mathbf{B}_0(t)\cdot \mathbf{S}_{\rm I} \otimes \hat{1} + 
\hat{1} \otimes \gamma_{\rm II} \mathbf{B}_0(t)\cdot \mathbf{S}_{\rm II}
\nonumber\\
& & + \frac{2J}{\hbar} S_{{\rm I};z} \otimes S_{{\rm II};z}
\end{eqnarray}
with ${\rm I}$ and ${\rm II}$ denoting the two qubits. 
By assuming $\gamma_{\rm I}/\gamma_{\rm II} \gg 1$, the Hamiltonian reduces to 
\begin{eqnarray}
H_0(t) & = & \gamma_{\rm I}\mathbf{B}_0^{(0)}(t)\cdot \mathbf{S}_{\rm I}\otimes \ket{0}\bra{0} 
\nonumber\\
 & & +\gamma_{\rm I}\mathbf{B}_0^{(1)}(t)\cdot \mathbf{S}_{\rm I}\otimes \ket{1}\bra{1} .
\label{eq:2roothamiltonian}
\end{eqnarray}
Now, we shall assume the rf field is swept to resonance, yielding $\gamma_{\rm I} 
\textbf{B}_0^{(q)}(t) \equiv \omega_{\rm I} (\cos\omega t , \sin\omega t , (1-2q) J/ 
\omega_{\rm I})$ with $q=0,1$ indexing the state of qubit ${\rm II}$ and we have defined 
$\omega_{\rm I} = \gamma_{\rm I} B_0$. Thus, ${\rm I}$ plays 
the role of the target qubit that sees two different effective magnetic fields $\textbf{B}_0^{(q)}(t)$ 
conditioned on the state of the control qubit ${\rm II}$. These effective fields are modified in the 
TQD scheme. If we define
\begin{eqnarray}
\cos\theta_q \equiv (1-2q) \frac{J}{\sqrt{\omega_{\rm I}^2+J^2}} \equiv (1-2q) \cos \tilde{\theta},
\label{eq:tildeangle}
\end{eqnarray}
the direction of each modified magnetic field is 
$\textbf{b}_0^{(q)}(t) = 
(\cos\omega t\sin\tilde{\theta},\sin\omega t\sin\tilde{\theta},(1-2q)\cos\tilde{\theta})$. 
The correction to the magnetic field in order to achieve the transitionless driving is then 
\begin{eqnarray}
\mathbf{b}_0^{(q)} \times \partial_t\mathbf{b}_0^{(q)} & = & 
-(1-2q) \omega \sin\tilde{\theta}\cos\tilde{\theta}
\nonumber \\ 
 & & \times (\mathbf{e}_x\cos\omega t + 
\mathbf{e}_y \sin\omega t) 
\nonumber\\
 & & + \omega \sin^2\tilde{\theta} \mathbf{e}_z , 
\end{eqnarray}
which define the full TQD effective fields 
\begin{eqnarray}
\gamma_{\rm I} {\bf B}^{(q)} (t) & = & \gamma_{\rm I} {\bf B}_0^{(q)} (t) + \mathbf{b}_0^{(q)} \times 
\partial_t\mathbf{b}_0^{(q)}  
\nonumber \\ 
 & = & \left[ \omega_{\rm I} - (1-2q) \omega \sin\tilde{\theta}\cos\tilde{\theta} \right] 
\nonumber \\ 
 & & \times (\mathbf{e}_x\cos\omega t + 
\mathbf{e}_y \sin\omega t) 
\nonumber \\ 
 & & + \left[ (1-2q) J + \omega \sin^2\tilde{\theta} \right] 
 \mathbf{e}_z 
\label{eq:2qubit_tqd}
\end{eqnarray}

The root Hamiltonian in Eq.~(\ref{eq:2roothamiltonian}) is 
diagonalised by its instantaneous eigenvectors
\begin{eqnarray}
\ket{\phi_{00}(t)} & = & \cos\frac{\tilde{\theta}}{2}\ket{00} + 
e^{i\omega t}\sin\frac{\tilde{\theta}}{2}\ket{10},
\nonumber \\
\ket{\phi_{10}(t)} & = & -\sin\frac{\tilde{\theta}}{2}\ket{00} + 
e^{i\omega t} \cos\frac{\tilde{\theta}}{2}\ket{10},
\nonumber \\
\ket{\phi_{01}(t)} & = & \sin\frac{\tilde{\theta}}{2} \ket{01} + 
e^{i\omega t} \cos\frac{\tilde{\theta}}{2}\ket{11},
\nonumber\\
\ket{\phi_{11}(t)} & = & -\cos\frac{\tilde{\theta}}{2}\ket{01} + 
e^{i\omega t}\sin\frac{\tilde{\theta}}{2}\ket{11} ,
\end{eqnarray}
where we have used that $\theta_0 = \tilde{\theta}$ and $\theta_1 = \pi - \tilde{\theta}$. 
The refocusing scheme extended to two qubits is 
\begin{eqnarray}
C \to \pi_{\rm I} \to \bar{C} \to \pi_{\rm II} \to C \to \pi_{\rm I} \to \bar{C} \to \pi_{\rm II} ,
\end{eqnarray}
where the Berry phases picked up are $\beta_{pq} = (2p-1)\frac{1}{2}\Omega_q$
with 
\begin{eqnarray}
\left| \Omega_q \right| & = & 2\pi \left[ 1-(1-2q) \cos \tilde{\theta} \right]  
\nonumber \\ 
 & = & 2\pi \left[ 1-(1-2q) \frac{J}{\sqrt{\omega_{\rm I}^2 + J^2}} \right] . 
\end{eqnarray}
Again, the solid angle changes sign depending on the 
direction of the path, thus $\tilde{\beta}_{pq} = -\beta_{pq}$, while $\tilde{\delta}_{pq} = 
\delta_{pq}$. The spin echo scheme results in
\begin{eqnarray}
\ket{\phi_{pq}} \to 
e^{i(\beta_{pq}-\beta_{p\oplus 1, q}+\beta_{p\oplus 1, q\oplus 1}-\beta_{p, q\oplus 1})}\ket{\phi_{pq}} .
\end{eqnarray}
Thus, the refocusing scheme enacts a purely geometric gate $\ket{\phi_{pq}} \to 
U(\Delta\Omega)\ket{\phi_{pq}} = e^{(-1)^{p+q}2i\Delta\Omega}\ket{\phi_{pq}}$, i.e.,
\begin{eqnarray}
U & = & e^{2i\Delta\Omega}(\ket{\phi_{00}}\bra{\phi_{00}}+\ket{\phi_{11}}\bra{\phi_{11}})
\nonumber\\
 & & + e^{-2i\Delta\Omega}(\ket{\phi_{01}}\bra{\phi_{01}}+\ket{\phi_{10}}\bra{\phi_{10}}) , 
\end{eqnarray}
where we have defined  the differential solid angle $\Delta\Omega \equiv (\Omega_1 - 
\Omega_0)/2$. Again, the prerequisite rotation of the magnetic field is required so that the 
substitution $\frac{\pi}{2} \to \vartheta$ can be made. This makes the unitary in the computational 
basis parameter dependent as $U(\vartheta_0,\vartheta_1,\Delta\Omega)$. By choosing 
$\vartheta_0 = \vartheta_1 = 0$ the phase gate 
\begin{eqnarray}
U(0,0,\Delta\Omega) & = & e^{2i\Delta\Omega}(\ket{00}\bra{00}+\ket{11}\bra{11}) 
\nonumber\\
 & & + e^{-2i\Delta\Omega}(\ket{01}\bra{01}+\ket{10}\bra{10}),
\end{eqnarray}
is obtained. This is a conditional gate where a phase difference of $4\Delta\Omega$ is picked 
up depending on whether the qubits are parallel  or not.

Since the TQD correction $\mathbf{b}_0^{(q)} \times \partial_t\mathbf{b}_0^{(q)}$ of the 
effective magnetic field ${\bf B}_q (t)$ seen by the target qubit ${\rm I}$ is 
conditionalized upon the state of the control qubit ${\rm II}$, it becomes difficult to realize 
experimentally the TQD Hamiltonian directly. This can be resolved by reformulating 
the system in terms of the time independent Hamiltonian 
\begin{eqnarray}
H_{\rm exp} & = & \omega_{\rm I}' \sin \theta' S_x \otimes \hat{1} + \omega_{\rm I}' 
\cos \theta' S_z \otimes \hat{1} 
\nonumber \\ 
 & & + \frac{2J_{zz}}{\hbar} S_z \otimes S_z + \frac{2J_{xz}}{\hbar} S_x \otimes S_z .
\end{eqnarray}
as the starting point for the scheme. In this way, the TQD Hamiltonian can be simulated by 
rotating the sample described by $H_{\rm exp}$ with angular frequency $\omega$ around 
the $z$ axis. This can be seen by evaluating the Hamiltonian in the rotating frame, yielding  
\begin{eqnarray}
\tilde{H}_{\rm exp} (t) & = & \gamma_{\rm I} \tilde{{\bf B}}_{\rm exp}^{(0)} (t) \cdot {\bf S} 
\otimes \ket{0} \bra{0} 
\nonumber \\  
 & & + \gamma_{\rm I} \tilde{{\bf B}}_{\rm exp}^{(1)} (t) \cdot  {\bf S} \otimes \ket{1} \bra{1}  + 
\omega \hat{1} \otimes S_z 
\label{eq:rotham}
\end{eqnarray}
with 
\begin{eqnarray}
\gamma_{\rm I} \tilde{{\bf B}}_{\rm exp}^{(q)} & = & 
\left[ \omega_{\rm I}'  \sin \theta' + (1-2q) J_{xz} \right] 
\left( {\bf e}_x \cos \omega t + {\bf e}_y \sin \omega t \right) 
\nonumber \\ 
 & & + \left[ \omega_{\rm I}' \cos \theta' + \omega + (1-2q) J_{zz} \right] {\bf e}_z . 
\label{eq:2qubit_exp}
\end{eqnarray}
Now, the extra control qubit term $\omega \hat{1} \otimes S_z$ in Eq.~(\ref{eq:rotham}) 
commutes with the Hamitonian $\tilde{H}_{\rm exp} (t)$ and will therefore be cancelled by the 
spin echo. We may thus simulate the effect of the two-qubit TQD system by a suitable 
choice of experimental parameters so that Eqs.~(\ref{eq:2qubit_exp}) and (\ref{eq:2qubit_tqd}) 
coincide. This yields  
\begin{eqnarray}
J_{xz} & = & - \omega \sin\tilde{\theta}\cos\tilde{\theta} = 
- \frac{\omega \omega_{\rm I}}{\omega_{\rm I}^2 + J^2} J , 
\nonumber \\ 
J_{zz} & = & J, 
\nonumber \\ 
\tan \theta'  & = & - \frac{\omega_{\rm I}}{\omega \cos^2 \tilde{\theta}} = 
- \frac{\omega_{\rm I}}{\omega J^2} \left( \omega_{\rm I}^2 + J^2 \right), 
\nonumber \\ 
\omega_{\rm I}' & = & \sqrt{\omega_{\rm I}^2 + \omega^2 \cos^4 \tilde{\theta}} 
\nonumber \\ 
 & = & \frac{\sqrt{\omega_{\rm I}^2 (\omega_{\rm I}^2 + J^2)^2 + 
 \omega^2 J^4}}{\omega_{\rm I}^2 + J^2} . 
\end{eqnarray}

Other high-speed geometric gates have been developed \cite{zhu02,xiangbin01,shao07,ericsson08} 
in the past. These gates share with our proposed one- and two-qubit gates that they are all based on 
the exact unitary evolution of quantum states. On the other hand, while these earlier proposals 
use nonadiabatic geometric phase concepts, such as the Aharonov-Anandan \cite{aharonov87} 
and Manini-Pistolesi \cite{manini00} phases, the geometric nature of our gates follows from its 
explicit relation to the adiabatic Berry phase of the root Hamiltonian $H_0 (t)$. In this way, our 
scheme combines the high-speed nature of nonadiabatic geometric quantum computation 
\cite{zhu02,xiangbin01,shao07,ericsson08} with parametric control of its adiabatic 
counterpart \cite{jones00,ekert00}.

\section{Conclusions}
A technique for accurate spin control based on a combination of spin echo with transitionless 
quantum driving has been proposed. This provides means to perform all elements in the 
spin echo sequence at high speed. We have demonstrated that the technique can be 
used to implement robust one- and two-qubit geometric gates in NMR quantum information 
processing. This form of high-speed geometric gates differs from previous proposals of 
nonadiabatic geometric quantum computation. Another potentially important technological 
application of the  high-speed spin echo scheme is MRI, for which the suppression of 
nonadiabatic transitions may help to improve scanning resolution. 

\section*{ACKNOWLEDGMENTS}
E.S. acknowledges support from the Swedish Research Council (VR) Grant No. 2017-03832.

\end{document}